\documentclass[
 aps, pra,
 amsmath,amssymb,
 11pt,
 final,
tightenlines,
 twoside,
 twocolumn,
 nofloats,
nofootinbib,
 superscriptaddress,
showkeys,
showkeywords,
onecolumn
 ]
{revtex4-2}

\usepackage[T1]{fontenc}
\usepackage[utf8x]{inputenc}
\usepackage[english]{babel}
\usepackage{graphicx}
\usepackage{dcolumn}
\usepackage{bm}
\usepackage{longtable}

\input{maik.rty}

\setcitestyle{authoryear,round}
\def\saoname{Special Astrophysical Observatory,  Russian Academy of Sciences,
              Nizhnii Arkhyz, 369167 Russia}

\def\squareforqed{\hbox{\rlap{$\sqcap$}$\sqcup$}}

\def\sq{\ifmmode\squareforqed\else{\unskip\nobreak\hfil
\penalty50\hskip1em\null\nobreak\hfil\squareforqed
\parfillskip=0pt\finalhyphendemerits=0\endgraf}\fi}

\def\utw{\smash{\rlap{\lower5pt\hbox{$\sim$}}}}

\def\udtw{\smash{\rlap{\lower6pt\hbox{$\approx$}}}}

\def\diameter{{\ifmmode\mathchoice
{\ooalign{\hfil\hbox{$\displaystyle/$}\hfil\crcr
{\hbox{$\displaystyle\mathchar"20D$}}}}
{\ooalign{\hfil\hbox{$\textstyle/$}\hfil\crcr
{\hbox{$\textstyle\mathchar"20D$}}}}
{\ooalign{\hfil\hbox{$\scriptstyle/$}\hfil\crcr
{\hbox{$\scriptstyle\mathchar"20D$}}}}
{\ooalign{\hfil\hbox{$\scriptscriptstyle/$}\hfil\crcr
{\hbox{$\scriptscriptstyle\mathchar"20D$}}}}
\else{\ooalign{\hfil/\hfil\crcr\mathhexbox20D}}%
\fi}}

\newcommand{\ab}{Astrophysical Bulletin }

\newcommand{\aap}{Astron. and Astrophys. }
\newcommand{\aas}{Astron. and Astrophys. Suppl. }

\newcommand{\aj}{Astron.~J. }

\newcommand{\mnras}{Monthly Notices Royal Astron. Soc. }

\newcommand{\pasp}{Publ. Astron. Soc. Pacific }
\newcommand{\arep}{Astronomy Reports }

\begin{document}

\selectlanguage{english}

\keywords{methods: data analysis; techniques: imaging spectroscopy, radial velocities; software:
development, public release}

\title{{\tt DECH}: Software Package for Astronomical Spectral Data Processing and Analysis}

\author{\firstname{Gazinur A.}~\surname{Galazutdinov}}
\email{runizag@gmail.com}
\affiliation{Federal State Budget Scientific Institution Crimean Astrophysical Observatory of RAS, Nauchny 298409, Crimea}
\affiliation{\saoname}

\begin{abstract}
The article provides a brief description of the software package {\tt DECH} for processing and analysis
of astronomical spectra. {\tt DECH} supports all stages of processing and analysis of spectral data, including
image preprocessing, spectra extraction (including those with a variable tilted slit), wavelength calibration
by a two-dimensional polynomial, continuum normalization (manual or automatic), measurement of
equivalent widths and radial velocities in various ways, cross-correlation analysis, etc. The {\tt DECH} software
package is actively used by astronomers from different countries and continues to be improved. In particular,
utilities for processing and analysis of data from the high-resolution fiber-feed echelle spectrograph
installed at 6-m telescope of Special Astrophysical Observatory of Russian Academy of Sciences were
added in the latest version. Software provides high-precision measurements of radial velocities, including
those for detection of extraterrestrial planets.
\end{abstract}

\maketitle

\section{INTRODUCTION}

The first version of the software package {\tt DECH}\footnote{1The program that can be freely distributed, available
for download http://www.gazinur.com/Download.html. The installation instructions—http://www.gazinur.com/installation.html}
was created in the early 90s of the last century, when the digital radiation detectors such as panoramic
photon counters, charge-coupled devices (CCDs) with all their advantages and disadvantages began
to be actively introduced into astronomical spectroscopy. {\tt DECH} is a highly specialized tool for solving
problems of astronomical spectroscopy, in contrast to universal systems for processing astronomical data,
such as, for example, {\tt IRAF} (Tody 1986) or MIDAS (Banse et al. 1983).

The software package {\tt DECH} is the Microsoft Windows OS—oriented, which is a significant difference
between {\tt DECH} and other popular programs. However, using MS Windows emulators you can bypass this
limitation. In particular, in the environment of such emulators, {\tt DECH} is fully functional in operating systems
Linux and Mac OS. The {\tt DECH} package was created in frame of the concept "an astronomer writes a
program for astronomers"\, which allowed to achieve a sufficiently high level of interface friendliness: mainly,
a work in the command line takes place only during the prepossessing of astronomical images, before the
spectra extraction. The extraction and subsequent analysis of the spectra is carried out in the “desktop”
mode, when all necessary tools are collected in one place and in the frames of one or two large integrator
programs, almost without using the command line. In contrast to fully automatic processing programs
(the so-called pipeline), we have abandoned the “black box” concept, in which all intermediate
data remain outside of user control (Fig. 1).

\begin{figure*}
\includegraphics[width=12.5cm]{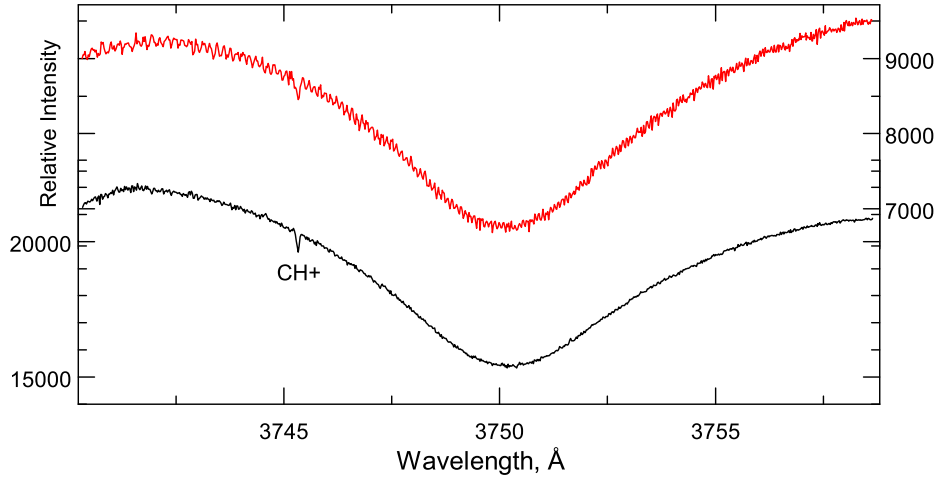}
\caption{Comparison of processing results obtained using the pipeline UVES spectrograph (top) with those obtained using
DECH (below). The sawtooth structure which is seen in the first spectrum probably resulted from an erroneous determination of
the order limits in the direction across the dispersion.} 
\label{Galazutdinov_fig1}
\end{figure*}

{\tt DECH} has become a popular tool among astronomers in the countries of former Soviet Union and abroad for more than 20 years of use. 
The development of a software package continues both to correct the identified errors and to solve the new astrophysical problems. 
This paper describes the main functions of {\tt DECH}, with a more detailed description of the processing data obtained using a new 
fiber-feed echelle spectrograph for the 6-m BTA telescope of the Special Astrophysical Observatory (Valyavin et al. 2014).

\section{OBSERVATIONAL DATA}

The spectral image—a table with the number of rows and columns which are corresponding or a multiple
(if we use software pixel binning) to the physical CCD matrix dimension is the result of observations
on modern echelle spectrographs equipped with CCD-type light detectors. Each CCD matrix
cell (pixel) works as an independent radiation detector. Properties of CCD-type light detectors such
as heterogeneity of pixel sensitivity, light interference on thin films of the device, etc, make it necessary
to obtain additional calibration data: bias, an image taken on a CCD without actual exposure time
and containing only an unwanted signal from the electronics that process the sensor data; flat-field, a
light source with a continuum spectrum, i.e. without spectral lines; comparison spectrum for building
a wavelength scale—the emission spectrum of a thorium-argon lamp (ThAr) is usually used in high-resolution
spectroscopy.

The potentially high quality of the observational material may be lost due to insufficient or poor quality
of calibration data. A few recommendations that will allow you to achieve optimal results, including in
unstable weather conditions:

\begin{list}{$\bullet$}
{
\setlength\leftmargin{6mm} \setlength\topsep{1mm} \setlength\parsep{0mm} \setlength\itemsep{1pt} }
\item 
bias—images without exposure (in fact, very short exposure values are
used, for example, 0.1 sec), without illumination of the light detector. The number of bias images
is determined by the level of the desired signal-to-noise (S/N) ratio. As a rule, with stable operation of
the CCD detector, the average of 10–20 images provides sufficient quality of the average bias
image. However, for tasks that require a very high signal-to-noise ratio (800–1000 or more),
the number of bias images should be increased to 100–150 (depending on the parameters of the
light detector). The {\tt ZEROC} procedure is used to obtain an average of several images with approximately
the same signal level. The procedure is performed on the command line, the original
images are cleaned using a median filter, after which the average value of the cleaned images is calculated;;
\item 
Flat field spectrum (flat field) — a continuum like spectrum of a laboratory light source, normally
the so-called hollow cathode lamps. The averaged spectrum of the flat field should have a S/N ratio
at least not lower than that of the observed objects over the entire wavelength range. Particular
attention should be paid to the blue part of the spectrum, where the efficiency of a laboratory
light source normally is relatively low. Usually, at least 10 flat field images are required to obtain an average with
a sufficient S/N level over the entire range, but if the S/N should be about 800–1000 or higher, the
required number of exposures can reach 100 or more. Dividing of the original spectra (or images)
by an averaged flat-field makes it possible to get rid of the effects of inhomogeneity in the sensitivity of
individual pixels and interference effects (fringes), which are especially strong in the red wavelength region of 
spectrum;
\item 
The optimal exposure time. As a rule, the special online calculators are used to determine the exposure
time of astronomical objects, e.g. UVES ETC\footnote{\url{https://www.eso.org/observing/etc/bin/gen/form?~INS.NAME=UVES+INS.MODE=spectro}}. However, the
exposure time calculated using such a calculator is often far from optimal, both due to inaccurate
initial data and weather–related reasons. The following approach is recommended: first of all
you should get a minute or second exposure depending on the brightness of the subject. Use
the CROSSCUT procedure (performed on the command line) to estimate the level of integration
achieved with a short exposure, then calculate the optimal exposure time necessary to achieve roughly
70 percent of the maximum possible integration level of the light detector in use. The maximum value
is usually around 65000, in which case a value around 40000 – 45000 can be considered optimal.
Do not forget to subtract the level of the bias during assessing the integration level of a
short exposure;
\item    
Cosmic particles. The sensitivity of CCD detectors to cosmic particles makes it impossible to 
obtain arbitrarily very long exposures. For efficient
cleaning of images from traces of the cosmic particles, it is recommended to observe each object at 
least twice with the same exposure. The duration of each exposure should not exceed 45–60 minutes;
\item 
Telluric lines. To solve some scientific problems, it becomes necessary to investigate the spectral
lines which are located in the region of strong telluric lines (see for example Galazutdinov et al.
(2017)). There are two ways to remove telluric lines: either using specialized programs that simulate
absorption lines originated in the Earth’s atmosphere (for example, {\tt Molecfit} (Smette et al. 2015),
{\tt TelFit} (Gullikson et al. 2014) etc.), or using the telluric standard ("divisor") spectrum of a hot
star without interstellar reddening (for example HD116658 (Spica) or HD120315), preferably
with high rotation velocity. Unfortunately, as a rule, the model spectra of telluric lines, do not always provide an
acceptable quality of removal of saturated telluric lines. The model method does not give an acceptable
result if the S/N ratio is more than 200. Dividing by the spectrum of the telluric standard
can provide a high quality of the cleaned spectrum if the following conditions are observed: the spectrum of
the object under study and the spectrum of the standard were obtained at approximately the
same zenith distances; the spectrum of the standard does not have lines with a complex profile
in the wavelength region of interest. Obviously, the spectrum of the standard should have a high
S/N—preferably at least 20 percent higher than that of the object under study. The procedure for removing
telluric lines is presented in the {\tt DECH-FITS} 
module of the program package\footnote{3option "Curve | Remove Telluric Lines"\, in the program
DECH95}. The procedure performs correction (if necessary) of the intensity of telluric lines and correction of their position
if the spectra are slightly shifted relative to each other. Examples of the spectrum before and after
cleaning can be seen in e.g. Galazutdinov et al. (2017).
\item 
Spectrum for the wavelength calibration — a comparison
spectrum. Usually, this is the spectrum of a thorium-argon lamp (ThAr). There are
very strong lines of argon in the red spectral range. As a rule, they are saturated, so that even
neighboring spectral orders can be affected. To avoid this effect, it is recommended to acquire
20 or more ThAr spectra with a relatively short exposure, and then to create an averaged image
using the {\tt ZEROC] procedure. This will make it possible to avoid the excessive influence of strong
argon lines on neighboring orders, to achieve a good S/N ratio even for weak thorium lines, and
at the same time to clear the spectrum of traces of cosmic particles. One can also successfully use
the Solar spectrum to control the wavelength scale. To do this, one can, for example, carry out the
spectrum of the reflected sunlight from a bright cloud or the Moon. The exact wavelengths of the solar
spectrum lines are given by, e.g. Allende Prieto and Garcia Lopez (1998).
}
\end{list}

\section{IMAGE PREPROCESSING}

Data preprocessing involves obtaining the average of the bias, the flat field, and the comparison spectrum
images, and then subtracting the first from all other images. To obtain an average image of any type, the
procedure {\tt ZEROC} is used. To subtract the average image bias from all the others, the procedure SUB
should be used. The procedure {\tt CROSSCUT} is used for preliminary control of the initial data homogeneity.

An important note regarding the flat-fielding: the fact is that the correction of the spectra
of astronomical objects using the flat-field spectrum can be performed in two ways—dividing the object
images by a flat-field image during image processing or by dividing of the extracted spectra after the
spectral extraction stage. The first method should be used for slit spectrographs. This is due to the
variable distribution of the signal over the height of the slit from object to object. A typical example—
UVES or MIKE spectrographs, where the slit height is significant, and the width of the spectral orders (in
direction across the dispersion) varies depending on the object brightness, the quality of the image on the
sky, the tracking accuracy, and other factors. The second method is preferable for fiber-feed spectrographs
(including a high-resolution spectrograph at the 6-m telescope of the SAO RAS).

To take into account the flat-field in the first way, that is, before the spectra extraction, after subtracting
the bias from the individual flat field images, one should obtain the average flat-field image (using the {\tt ZEROC} program),
then subtract the scattered light from it. To do this, one will have to determine the position and width
of the spectral orders (see Section 4.1). The image must be normalized after subtracting the scattered 
light\footnote{Procedure "Mask|Normalize apertures (Apnormalize)"\, in {\tt DECH95} program}.
Dividing of the original images by the normalized flat field with the averaged bias already subtracted can be
performed by {\tt FLATC} or {\tt IMARITH} programs.

Fiber-feed spectrograph data preprocessing, including the FFOREST spectrograph at the 6-m BTA SAO RAS
telescope is performed in two stages. First, the average bias image have to be obtained for the subsequent subtracted of it from
all other images, after that the extraction procedure can be performed.

\begin{figure*}
\includegraphics[width=16cm]{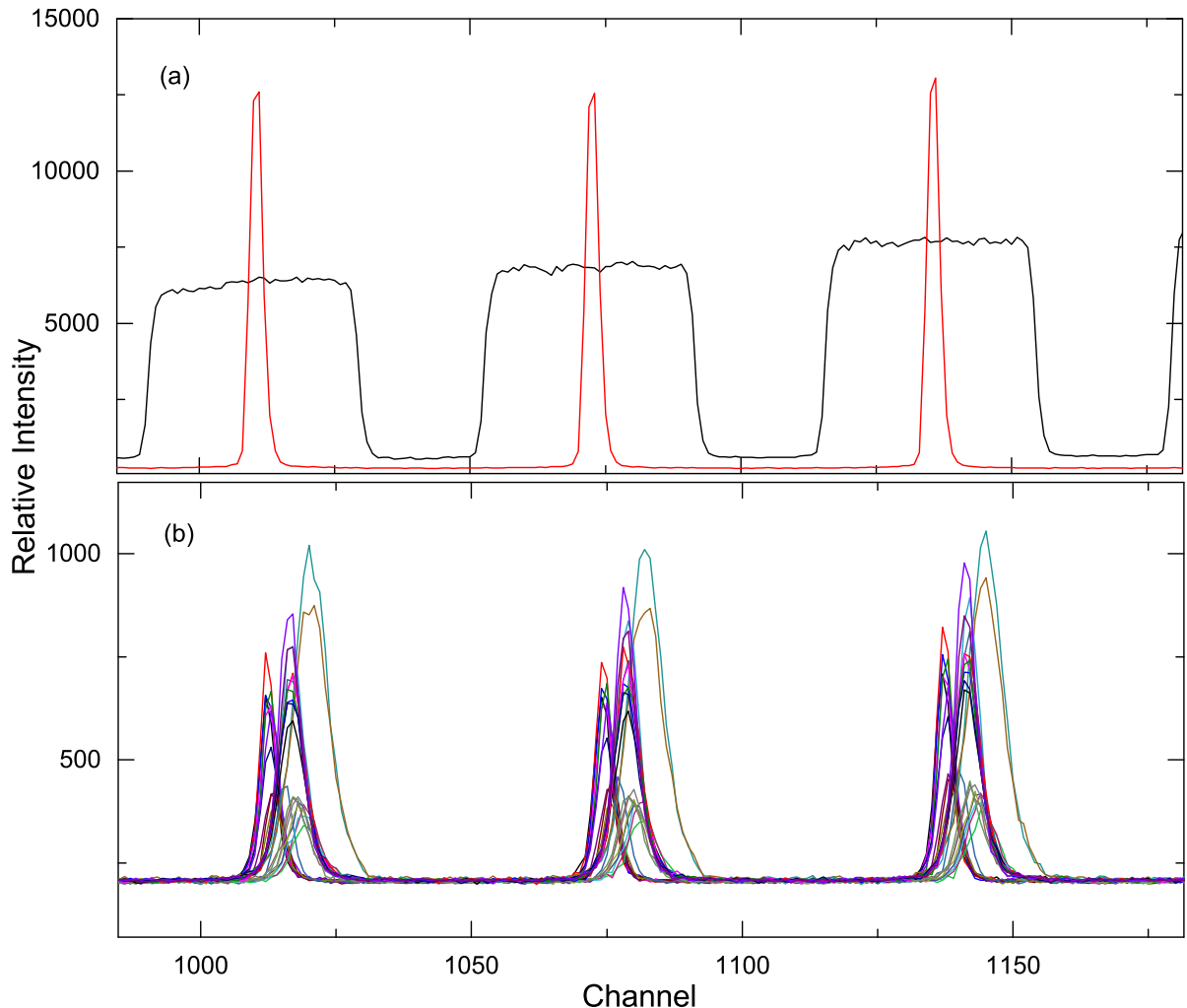}
\caption{Fragments of the transverse sections of UVES spectrograph images. {\bf (a)} – a flat field spectrum (wide orders) and special
image spectrum ORDERREF serving for finding the position of spectral orders. {\bf (b)} — a set of spectra of astronomical
objects obtained during one night with the same spectrograph configuration. The variability in the position and shape of the
spectral orders from object to object is clearly seen.
}
\label{Galazutdinov_fig2}
\end{figure*}

\section{EXTRACTION SPECTRA}

The {\tt DECH95} program is used to extract spectra from images. The procedure includes the following operations:
\begin{list}{$\bullet$}{
\setlength\leftmargin{8mm} \setlength\topsep{1mm} \setlength\parsep{0mm} \setlength\itemsep{1pt} }
  \item building a mask, that is, determining the position (trajectory along the main dispersion direction) and the boundaries of the spectral orders (width across the dispersion);
  \item scattered light subtraction;
  \item extraction.
\end{list}

\subsection{Building a mask}

The mask is a table of information about the position and width of the spectral orders. In {\tt DECH}
the mask is given as a set of Chebyshev polynomial coefficients for each spectral order separately. The
mask made using the {\tt DECH} is written in the IRAF processing system format and can be used for spectrum
extraction by IRAF tools, such as {\tt apall} or {\tt apsum}.

Fiber feed optic spectrographs are characterized by a high stability of the position of spectral orders;
therefore, as a rule, one mask is sufficient for all images which are obtained in a fixed instrument configuration.
The situation is more complicated for the slit spectrographs: both the position and the width of
the spectral orders can change from object to object for the reasons indicated above. Therefore, a mask correction
is needed for each individual image (Fig. 2b).

To construct the main mask, the best choice is spectral images of a flat field or special images carried out with the spectrograph's 
slit reduced in height, as it is done at the UVES spectrograph(Fig. 2a). Theoretically,
to build a mask, it is allowed to use a well-integrated image of any object with a pronounced continuum
spectrum, e.g. hot fast rotating main sequence OB stars. However, in this case, one may have problems with determining
the position of spectral orders in areas with saturated telluric lines (Fig. 3). This may cause the need for interactive mask correction.

\begin{figure*}
\includegraphics[width=14.5cm, bb=25 250 570 590,clip]{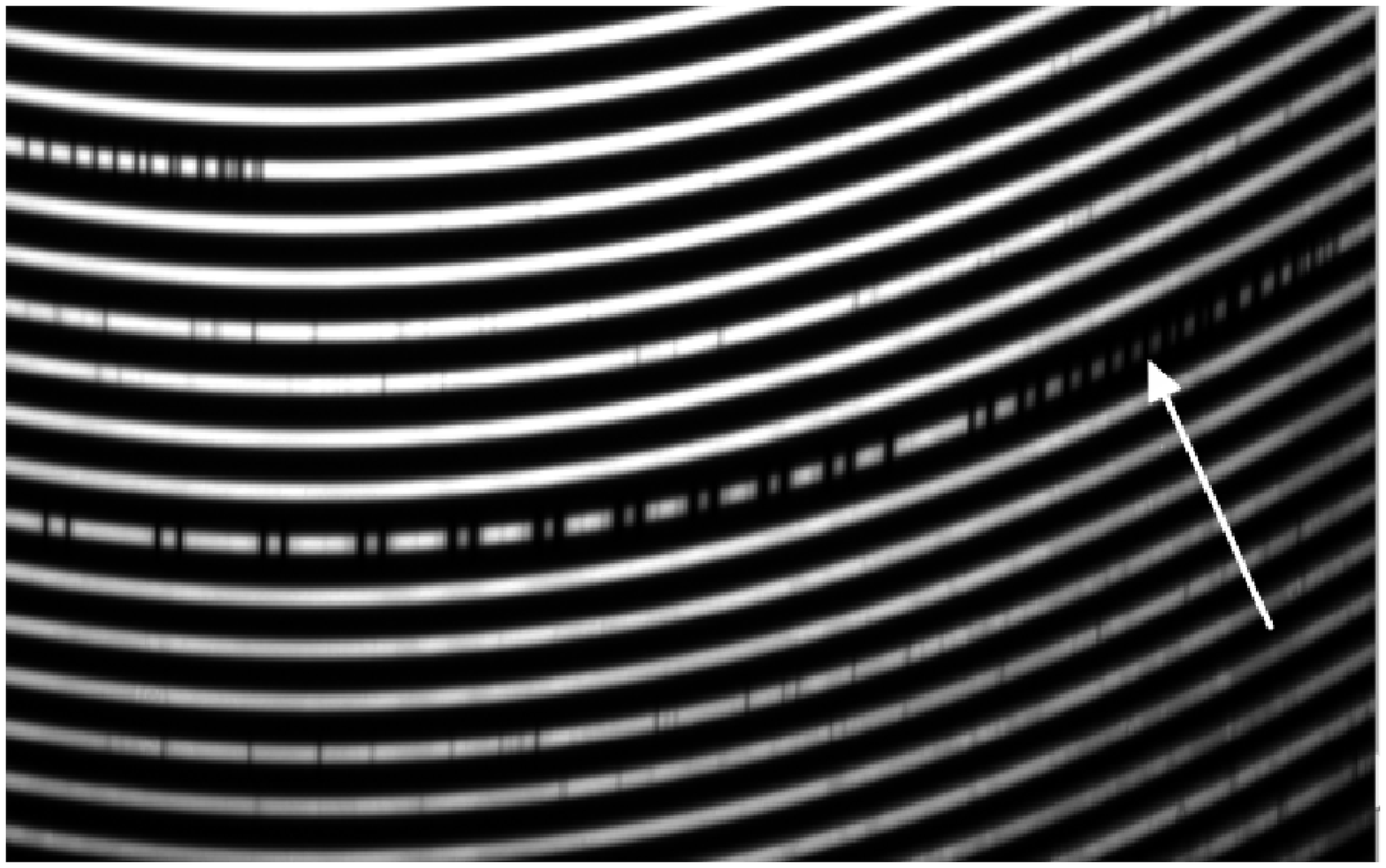}
\caption{A fragment of the spectral image of an astronomical object. The arrow indicates the area of saturated telluric lines.} \label{Galazutdinov_fig3}
\end{figure*}

A "preliminary"\, mask is a set of curves approximately passing trough the centers of the cross-sections
of the spectral orders over the entire orders length.
At the first stage, the orders' width is not determined.
The center of the spectral order is a quite conditional
term. It is not recommended to use a pixel with with maximum
signal as the center, in particular for spectra with
a low S/N. The cross-sectional profile
of the order is often not symmetrical. This is typical,
for example, for slit spectrographs or spectrographs
with an image slicer. Therefore, it is optimal to use
the center of gravity  as the center of the cross section of the order (as shown in equation 1). 
The equation defines the center of gravity of the spectral order $m$ at
the cross section through the pixels $x$. The summation
is carried out along the vertical (it is perpendicular to
the main dispersion) from the lower ($v1$) to the upper
limit of the order ($v2$)::

\begin{equation}
\label{gravcent} G_{\rm {mx}} = \displaystyle\sum_{y=v1}^{v2} I_{xy} \times y / \displaystyle\sum_{y=v1}^{v2} I_{xy},
\end{equation}
where $I_{xy}$ — the signal intensity in a pixel with coordinates
$x$ and $y$.

To create a "preliminary"\, mask it is necessary to determine the centers of gravity along each spectral
order with a certain step\footnote{Procedure "Mask |CreateMask"\, in the {\tt DECH95} program}. 
It is usually, every 10–20 pixels along the entire length of the order. Then, an
approximating curve is drawn through these centers for each order separately. As a rule, the Legendre or Chebyshev 
polynomial of 3–5 degrees is used. A spline approximation may be required in some rare cases.

\begin{figure*}
\includegraphics[width=16.cm]{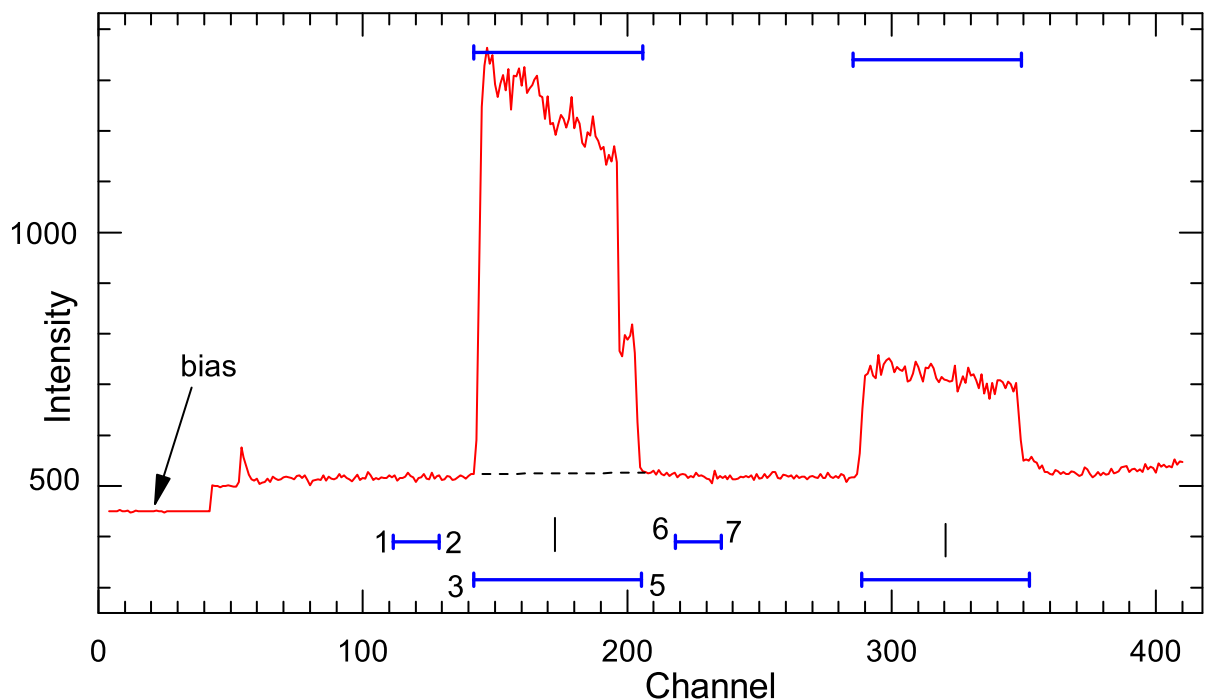}
\caption{A fragment of the cross section of the echelle spectrum with two spectral orders. Bias level is marked. The
boundaries of the left order (from 3 to 5) and the boundaries of two regions of inter-order minima to the left and right of the order
are shown (numbers 1, 2 and 6, 7, respectively). The dashed line shows the level of the scattered light.} \label{Galazutdinov_fig4}
\end{figure*}

The next step involves determining the boundaries of the orders and the boundaries of the inter-order minima (Fig. 4). 
These procedures are performed in mask editing mode\footnote{Procedure "Mask | Edit"\, in the {\tt DECH95}} . Incorrect definition of the order boundaries can lead to a sawtooth pattern or an unnecessarily low S/N ratio if the order width is insufficient or unreasonably increased. Incorrect determination of the boundaries of inter-order minima
leads to errors in the subtraction of scattered light with fatal consequences for the quality of the extracted
spectrum: it becomes impossible to correctly take into account the flat field; the intensities of the
spectral lines and, accordingly, the equivalent widths will be distorted.

Automatic determination of the order boundaries in {\tt DECH} is performed after the scattered
light level has been determined. To do this, an approximating curve is drawn through the zones of inter-order
minima in the mask editing mode. The type of the optimal approximating function is interactively
determined. As a rule a broken curve ($PolyLine$) gives a good result. After having determined the level of
scattered light, the option to automatically determine the boundaries of the orders becomes available. Automatic
order width detection may not be correct for the orders with low S/N ratio. Usually, the manual
correction of order boundaries and zones of inter-order minima is required for the first and last orders,
where the S/N ratio is the lowest. 

The resulting mask is saved in a special file in {\tt IRAF} format with
the possibility of subsequent changes. It is important to note that the type of functions approximating the
scattered light distribution is generally not related to the mask. This information is used in the extraction
step. The mask contains information only about the trajectory (shape) and the width of the spectral orders.

The scattered light is a two-dimensional function of a complex shape. To subtract the scattered light,
at the first stage “carrier” curves are determined as passing through the zones of inter-order minima in
the direction perpendicular to the main dispersion. Such curves are built along the entire image with an equal step, 
normally, every 10 pixels. At the next stage, fitting curves are drawn through these “carriers” passing in the direction 
of the main dispersion through each line of an image. Normally, spline interpolation with preliminary smoothing of the initial 
data by seven points gives a good result.

Each order on the spectrum image of the BTA fiber-feed spectrograph has two sub-orders: one suborder
serves to register an astronomical object, the second one to register the spectrum of a thorium-argon
lamp. The latter one serves to control the stability of the spectrograph, which is necessary
for accurate measurements of radial velocities. Therefore, to extract the spectra obtained using this
spectrograph, it is necessary to have two masks: the first one for sub-orders with an astronomical object
spectrum, the second for sub-orders with a comparison spectrum. It is not convenient to build
a mask, when we use the images with two visible sub-orders. As it was mentioned above, it is more
convenient to use the flat field spectrum without using the comparison spectrum channel. Such a mask,
constructed from the flat field spectrum, can be used for the extracting the astronomical objects spectra.
To create a mask for extracting the comparison spectrum,
it is enough to shift the "stellar"\, mask by a fixed number of pixels to the left (in the FFOREST configuration
for 2022) or to the right, depending on the arrangement of sub-orders in the images. This procedure is
performed in the mask-edit mode {\tt("Mask | Edit Mask")} in {\tt DECH95} program.

\subsection{Extraction}

The spectrum in the conventional sense as a result of extraction is a two
column table: the pixel number is in the first column and, the intensity — in the second one\footnote{Procedure "Extraction | Using mask...\, in the {\tt DECH95}}. 
During the further processing, the pixel number is converted into a wavelength.

The extraction procedure includes subtracting the scattered light, taking into account the tilt of the
slit (if it is necessary), and actually the spectrum extraction which is the signal summation in the direction
perpendicular to the direction of the main dispersion, that is, along the width of the spectral
order, within the boundaries defined by the mask and along the entire length of the order. Extraction is performed
separately for each spectral order, the result is saved in a {\tt FITS} format file. As mentioned above,
in some cases, for example, for {\tt UVES} spectra, the flat field is taken into account at the image processing
stage. In such a case, scattered light subtraction must be performed prior to extraction. For this case,
we use a mask. The sequence of actions is exactly the same as during extraction, with the difference that
the scattered light subtraction trigger is turned on in the extraction procedure. The subtracted images of
scattered light are saved in files with the "sub"\, suffix. 
In the same way, scattered light is subtracted from the averaged flat-field image.
Then the flat field image is normalized\footnote{Procedure "Mask | Normalize Apertures (Apnormalize)..."\,
in the {\tt DECH95}}. After that, all images with scattered light subtracted are divided by the normalized flat
field\footnote{Dividing the images can be performed in  a command line using {\tt FLATC} or {\tt IMARITH} programs}.

There are two main methods of spectra extraction from echelle images: simple integration and the
so-called optimal extraction (Horne, 1986). The summation is the simplest and fastest method. To
extract the spectrum in vector form, the image data is summed in each column (across the direction of the
main dispersion within order width) with equal weights at a given aperture:

\begin{equation}
\label{sum} F(x) = \displaystyle\sum_{y=v1}^{v2} I_{xy}.
\end{equation}

\begin{figure}
\includegraphics[width=8.2cm]{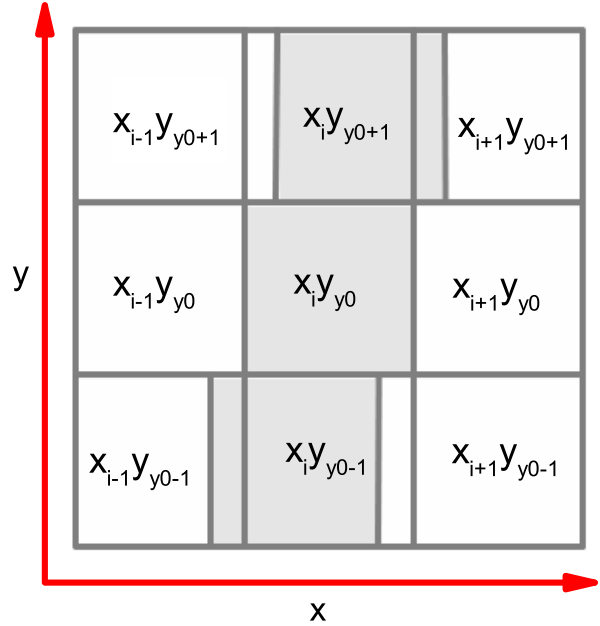}
\caption{An example of signal distribution on the CCD
matrix in the presence of a slit tilt. The gray color shows
the projection of the inclined slit onto the CCD array.} 
\label{Galazutdinov_fig5}
\end{figure}

The "optimal extraction"\, method was developed in the works of Horne (1986), Marsh (1989),
Mukai (1990). The essence of the method lies in the fact that when we integrate the cross section of the
spectrum, the weight of each pixel is preliminarily estimated: the lower the signal at the profile point,
the lower its weight. This makes it possible to reduce the effect of pixels on the periphery of the
cross-section profile. A more precise definition of the order bounds makes it possible to reduce the
importance of applying optimal extraction. Indeed, when we integrate the spectrum, we are faced with
a dilemma: on the one hand, it is desirable to have the integration aperture as wide as possible in order
to capture the entire useful signal, but, on the other
hand, at some distance, conditionally, far from the order center, the signal is already significantly noisy,
and if we take into account this signal it can lead to
a deterioration in resulting S/N ratio. To solve this problem, we
must define a $P(x, y)$ profile that reflects the signal
distribution in the order. An indispensable condition
is the normalization of the profile within the framework
of the integration aperture $\sum P(x,y)=1$.
The optimization itself consists of the subsequent
redefinition of $P(x, y)$: for each $y$ coordinate, the cut
is approximated by, e.g. a polynomial, and this process
is repeated, discarding points that deviate from the
fitting curve by more than $k\sigma$ ($k = 2\ldots3$), until
all used points will be within $k\sigma$. However, the
described optimal extraction scheme is really effective
only when we work with spectra with a sufficient
number of points in a profile, that varies slightly with
wavelength. The optimal extraction methods with
curved spectra and spectra with varying spatial profile
were developed in papers of Marsh (1989); Mukai
(1990): to determine the spatial profile $P(x, y)$ more
accurately, the concept of "virtual re-sampling"\, of the
spectrum is introduced: between spectrum profile
pixels (in the cross-dispersion direction) introduce
additional virtual “pixels” (usually no more than
10), the counts in which are redefined based on
linear regression so that the total signal remains
unchanged. If the profile of the spectrum varies
greatly with wavelength, the attempt to “smooth”
it may ruin the spectrum. In this case, the split of
the spectrum into fragments by wavelength, in which
the profile change is insignificant can be a simple
and effective solution. Both methods of spectrum
extraction are implemented in {\tt DECH}: the standard
method and optimal one. The optimal extraction can
give a positive effect only for the spectra with a low
signal-to-noise ratio, the conventional simple integration is
more preferable for well exposed spectra.

\subsubsection{Correction for slit inclination}

In some echelle spectrographs (for example, MIKE spectrogragh (Bernstein et al. (2003)
at 6.5-m telescope  Clay  in Las Campanas observatory) spectral lines in spectral images deviate significantly
from the perpendicularity to the main dispersion, and the degree of deviation varies both along orders and along the direction of the
main dispersion. The conventional spectrum extraction from such images leads to a loss of spectral resolution and
distortion of the spectral lines profile.

To estimate the slit inclination magnitude  we introduce the parameter $N_{xy} = \Delta x/\Delta y$ in such a
way that $N_{xy}=0$ if the slit is located strictly vertically, i.e. there is no tilt.
For example, if $N_{xy}$$\sim$0.2 (Fig. 5), spectrum integration (extraction) with the tilted slit correction should be performed as follows:
\begin{equation}
\begin{array}{rcl}
        &\!+\!\!&0.2 I(x_{i+1},y_{y0+1}) +\!  \ldots\! \\[-4pt]
        &\!+\!\!&0.8  I(x_i,y_{y0-1})\\[-4pt]
        &\!+\!\!&0.2 I(x_{i-1},y_{y0-1}) +\! \ldots \\[-4pt]
  \end{array}
\label{one}
\end{equation}

The complete algorithm looks like this:

\begin{list}{$\bullet$}{
\setlength\leftmargin{6mm} \setlength\topsep{1mm} \setlength\parsep{0mm} \setlength\itemsep{1pt} }
  \item Construct a  $\Delta x/\Delta y$ values map over the entire image. To build such a map, one should use the
thorium-argon spectrum with a sufficiently high slit;
  \item 
  Approximate the resulting tilt estimates by a two-dimensional polynomial, so that for any
coordinate  $(x,m)$  of the spectral image (where $x$ — pixel numbers, $m$ — order number), one can
calculate the $N_{xy}$ value. In the case of the MIKE spectrograph, the second degree of the polynomial
is sufficient for both coordinates\footnote{Procedure "Mask | Solve Tilt Surface"\, in {\tt DECH95} program}:
  \item 
  Apply the obtained resulting solution in the extraction extraction procedure\footnote{Option "Fix the Tilted Slit"\, ~ in the procedure "Extraction | Using mask"\, in {\tt DECH95} program}.
\end{list}

A comparison of results of extraction with and without the tilt correction shown in Fig. 6.

\begin{figure*}
\includegraphics[width=12cm, bb= 25 245 565 603,clip]{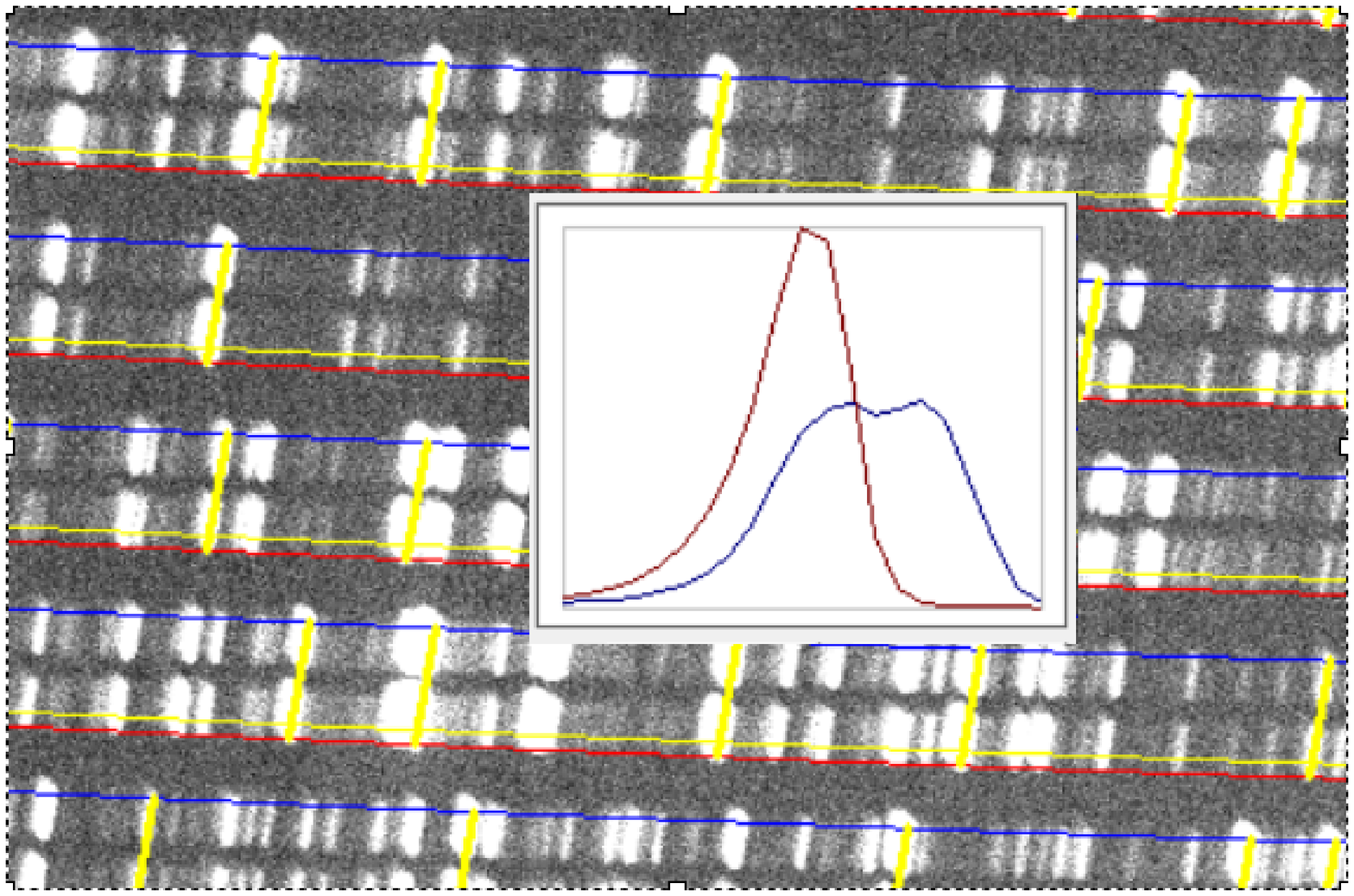}
\caption{A fragment of a ThAr echelle spectrum obtained using the MIKE spectrograph. A fragment
of the tilts map is visible (straight inclined lines passing through selected lines of thorium) built using the procedure "Mask |
Make TiltMap"\, of the {\tt DECH95} program. Two extracted profiles are shown: the integration result with taking into account the tilt of
the slit (red profile) and, the integration result without the correction (blue profile with two peaks).}
\label{Galazutdinov_fig6}
\end{figure*}

\section{WORK WITH THE EXTRACTED SPECTRA}

Extraction procedure for each FFOREST's spectral image produces two FITS
files: one file contains the actual spectrum extracted from the sub-order of the object, the
other one contains the corresponding thorium-argon spectrum extracted from the sub-order of the artificial
stars. The second file ends with "\_ThAr.fits"\,. The beginning of the file name coincides with the base file
name of the object spectrum.

To make the wavelength scale, it is recommended to use the averaged image of the thorium-argon spectra
with identical, rather short exposures (see Section 2).
The process of building the dispersion curve in the echelle spectrum is approximately the same as in
IRAF. First step — identification of thorium lines in adjacent or close two spectral 
orders\footnote{Procedure "Curve | Dispersion Curve Workshop | Create It"\, in the {\tt DECH-FITS} }, approximation of these identified knots
 by a two-dimensional polynomial\footnote{Procedure "Curve | Dispersion Curve Workshop | Global Fit \& Create FDS file"\, in {\tt DECH-FITS} program} in the form:

\medskip

   $\lambda(x,m) = \sum_{i=0}^{k} \sum_{j=0}^{n} a_{ij} x^{i} m^{j}$

\medskip
where $a_{ij}$ — the polynomial coefficients; $x$ — knots coordinate in the pixel space (along main dispersion);
$m$ — knots coordinate in the space of spectral orders. The first resulting solution made for two identified orders is not final, of course, 
but it allows us to identify the bulk of the lines in other spectral orders in the automatic mode. Sometimes it may
be necessary to manually identify the first and last orders to avoid the extrapolation, then repeat the
approximation and after that, identify all other lines using the automatic procedure. For the FFOREST spectrograph (as of 2022 year)
the recommended degree of the polynomial for the pixels space is 5 and, for the orders is 4.
Usually, the dispersion curve is calculated using about of 700–1000 knots. The means square
approximation error does not exceed 0.003 \AA. The resulting wavelength solution stored in the {\tt IRAF} format to
all spectra of the observing run, then the spectra are ready for further processing and analysis both in {\tt DECH} or 
{\tt IRAF}.

The continuum normalization can be done either manually or automatically. The automatic normalization
gives a good result for the spectra with narrow lines, e.g. in solar-type stars (see Fig. 7). In 
spectra with broad lines, exhibiting emissions and/or strong blending it is preferable to work in the manual mode.

\begin{figure*}
\includegraphics[width=12 cm, bb= 0 0 399 223,clip]{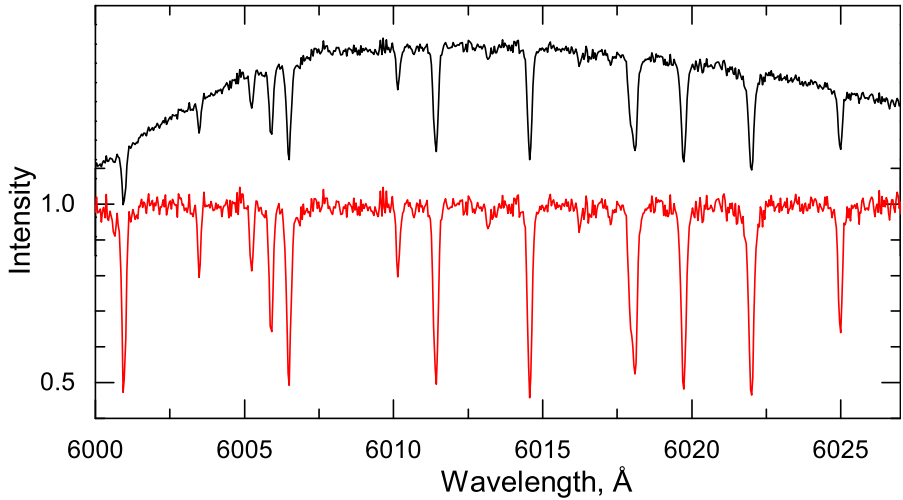}
\caption{An example of automatic continuum normalization.} \label{Galazutdinov_fig7}
\end{figure*}

\subsection{Measurement of the radial velocities by the cross-correlation method.}

The cross-correlation method is actively used in astronomy to measure the radial velocity of observed
objects. The method, which was proposed by Simkin (1974) and detailed in the paper of Tonry and
Davis (1979) was implemented in {\tt DECH} as well. To measure the shift of spectral lines in the spectrum of an object
observed on different dates two procedures are required: monitoring the stability of mechanical and optical part of the spectrograph; 
measurement of the shift of spectral lines in the investigated spectrum relative to the sample spectrum (template). The second procedure,
in turn, includes the actual measurement of the displacement and the calculation of the barycentric correction
based on the algorithm proposed by Stumpf (1979). The accuracy of determining the barycentric
correction is not worse than 50 ~sm\,s$^{-1}$ which places high demands on the accuracy of the initial data,
such as the effective exposure time, the coordinates of the observatory and of the observed object.

In the FFOREST spectrograph  the projections of the outputs of two fibers that form
spectra of neighboring sub-orders are located not strictly vertically under each other, but at a certain
angle conditioned by the considerable thickness of the fibers clad and the limitations on the space
between sub-orders. The presence of this tilt leads to the fact that the wavelength scale built on the suborder
of  the thorium-argon has a constant shift relative to the wavelength scale of the sub-order with the
spectrum of object. As of August 2022 year, the shift value is approximately 127 km\,s$^{-1}$.

As already noted, the control of the mechanical stability of the device is carried out using the thorium-argon
spectrum, which is observed simultaneously with the astronomical object. In Fig. 8 one can evaluate
the mechanical stability of the device on 19 consecutive exposures obtained within about 60 minutes
on September 26, 2021. The experiments with the equipment showed that the largest contribution to the
measurement error is made by spectral orders with one or two dominant lines, which are not necessarily
saturated. The best result was achieved with masking such lines. For cross-correlation, wavelength
areas free lacking the strong lines were chosen. As of August 2022, the measurement error does not exceed
10–20 ~m\,s$^{-1}$for stellar spectra and is approximately 1–2 ~m\,s$^{-1}$ for thorium-argon. With an increase in the
number and length of spectral orders after installing a permanent camera, the accuracy should increase at
least twice. The mechanical stability of the device will be significantly improved after the installation of a
continuous nitrogen supply system at the cryostat of the CCD camera.

\begin{figure*}
\includegraphics[width=14.5cm]{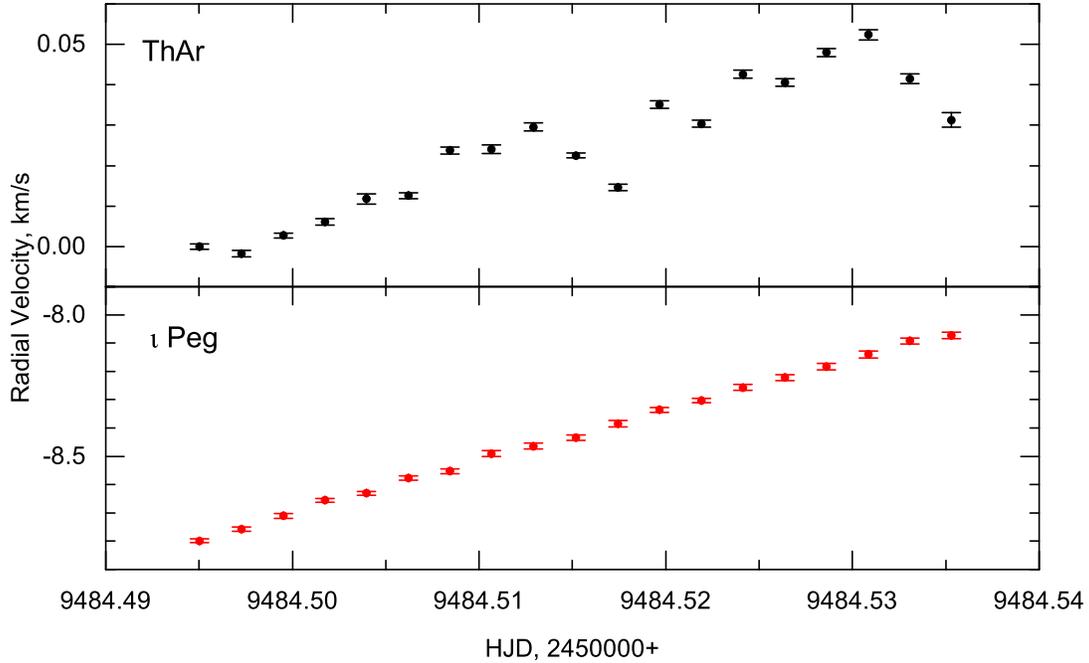}
\caption{Cross-correlation analysis of spectra of $\iota$\,Peg (lower plot {\bf(b)})
received within one hour of observations on September, 26,
2021. All corrections are taken into account, including the correction for the mechanical instability of the device {\bf (a)}, controlled
using simultaneous registration of thorium-argon spectrum.} 
\label{Galazutdinov_fig8}
\end{figure*}

To measure the radial velocity at any point in a selected line profile, {\tt DECH} provides an unique method to match the
direct and mirror line profiles. If to combine manually fragments of the mirrored (along the
wavelength axis) and the original profile under study, it is possible to estimate the radial velocity
of any component or part of a profile of any complexity. 

\subsection{Measuring equivalent widths}

\begin{figure*}[bpt!!!]
\includegraphics[width=14cm]{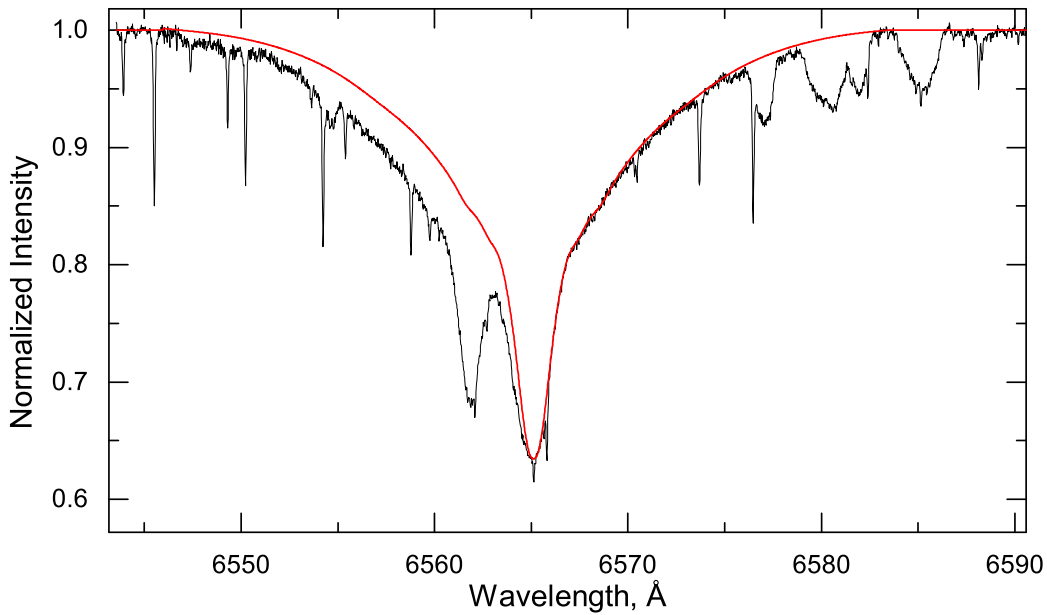}
\caption{An example of measuring the equivalent width with an arbitrary profile.} \label{Galazutdinov_fig9}
\end{figure*}

{\tt DECH} provides a possibility to measure equivalent widths (EW) in several ways. Classical methods
include direct integration, line fitting with Gaussian, Lorentzian, Voigt profiles, and automatic blend separation.
It is also possible to estimate the equivalent width by manually constructing a voluntary profile.
This method may be the only way to measure lines with strong blending, when the spectrum contains
only a (small) part of the profile of the line of interest (Fig. 9).

For the convenience of the measurements and results analysis, it is possible to use an identification list
of known lines in various formats, including data from the {\tt VALD} (Pakhomov et al. 2019) database. When such
lists are used, the measurement results are stored
together with all atomic data of the line, which simplifies
the subsequent analysis of the spectra by the stellar model
atmosphere  method. An error estimate in
measuring equivalent widths is not a trivial task and
has no general solution. In {\tt DECH} the EW-error is 
calculated in two ways. For strong and
wide lines, it is recommended to use the EW-error estimation 
based on the equation 7 from the paper of Vollmann
and Eversberg (2006) where, the measured line intensity is taken into account. For
weak and narrow lines, it is recommended to use the EW-error estimation calculated as the product of the line width
and the reciprocal of the signal-to-noise ratio\footnote{denoted as errOld in {\tt DECH-FITS} program}.

\section{CONCLUSION}

As of 2022, the high-resolution fiber-feed echelle
spectrograph installed at 6-m telescope of Special
Astrophysical Observatory of Russian Academy of
Sciences does not have a standard camera, so a
significant part of the spectrum at the edges of the
spectral orders is vignetted. As a result, a significant
part of the light detector of the CCD camera remains
unused, so it is recommended to pre-cut the informative
part of the observed image\footnote{One can trim an image fragment 
with the informative part using the command $trim 124 1633 266 1790 *.fits$. The
command must be executed for all images before all subsequent
processing steps. After installing the permanent spectrograph
camera, this step can be omitted}.

The {\tt DECH} software package has been used by astronomers
from the different countries for processing
and analyzing astronomical spectra for more than 20
years. The hundreds of publications have been written
based on measurements made with {\tt DECH}. The {\tt DECH} software package is under continuous improvement. 
The latest version includes utilities for 
analyzing data from the high-resolution fiber-feed echelle spectrograph {\tt FFOREST} at SAO RAS,
which makes it possible to perform high-precision measurements of radial velocities, 
and to detect exoplanets. A detailed description of the procedures and commands is given the user manual of the spectrograph.

\section*{FUNDING}
The project was supported by the Ministry of Science
and Higher Education of the Russian Federation,
grant No. 075-15-2020-780.

\section*{CONFLICT OF INTEREST}
The authors declare no conflicts of interest.

\bibliographystyle{aspb1}
\bibliography{DECH}

\begin{thebibliography}{25}
\providecommand{\natexlab}[1]{#1}
\bibitem[{Banse} et al.(1983)] {Banse83} K.~{Banse}, P.~{Crane}, P.~{Grosbol} et al., The Messenger \textbf{31}, 26 (1983).
\bibitem[{Bernstein} et al.(2003)] {Bern2003} R.~{Bernstein}, S.~A.~{Shectman}, S.~M.~{Gunnels}, et al.,  SPIE \textbf{4841}, 1694 (2003).
\bibitem[{Allende Prieto} and {Garcia Lopez}(1998)]{Prieto98} C.~{Allende Prieto},  R.~J.~{Garcia Lopez}, \aas \textbf{131}, 431 (1998).
\bibitem[{Galazutdinov} et~al. (2017)]{Galaz2017} G.~A.~{Galazutdinov}, V.~V.~{Shimansky}, A.~{Bondar}, et al., \mnras \textbf{465}, 3956 (2017).
\bibitem[{Gullikson} et~al.(2014)]{Gullikson2014} K.~{Gullikson}, S.~{Dodson-Robinson}, A.~{Kraus}, \aj \textbf{148}, 53 (2014). 
\bibitem[{Horne}(1986)]{Horne1986} K.~{Horne}, \pasp \textbf{98}, 609 (1986).
\bibitem[{Marsh}(1989)]{Marsh1989} T.~R.~{Marsh}, \pasp \textbf{101}, 1032 (1989).
\bibitem[{Mukai}(1990)]{Mukai1990} K.~{Mukai}, \pasp \textbf{102}, 183 (1990).
\bibitem[{Pakhomov et~al.}(2019)]{Pakhomov19} Yu.~V.~{Pakhomov}, T.~A.~{Ryabchikova} and N.~E.~{Piskunov}, \arep \textbf{63}, 1010 (2019).
\bibitem[{Simkin}(1974)]{Simkin1974} S.~M.~Simkin, \aap \textbf{31}, 129 (1974).
\bibitem[{Smette} et~al.(2015)]{Smette2015} A.~{Smette}, H.~{Sana}, et al., \aap \textbf{576}, A77 (2015).
\bibitem[{Stumpf} (1979)]{Stumpff1979} P.~{Stumpff}, \aas  \textbf{78}, 229 (1979).
\bibitem[{Tody}(1986)]{Tody1986} D.~{Tody}, SPIE \textbf{627}, 733 (1986).
\bibitem[{Tonry} and {Davis}(1979)]{TD1979} J.~{Tonry}, M.~{Davis}, \aj \textbf{84}, 1511 (1979).
\bibitem[{Valyavin} et~al.(2014)]{Valyavin2014} G.~G.~{Valyavin}, V.~D.~{Bychkov}, et al., \ab, \textbf{69}, 224 (2014).
\bibitem[Vollmann and Eversberg (2006)]{2006AN....327..862V} K.~Vollmann and T.~Eversberg, Astronomische Nachrichten, \textbf{327}~(9), 862 (2006).
\end{thebibliography}


\end{document}